\documentclass[final,superscriptaddress,english,twocolumn,amssymb,aps,prb,
longbibliography]{revtex4-1}
\usepackage{graphicx}
\usepackage{natbib}
\usepackage{amsmath}
\usepackage{amssymb}
\usepackage{appendix}
\usepackage{bm}
\usepackage{overpic}
\usepackage[dvipsnames]{xcolor}{\huge }
\usepackage[section]{placeins}
\definecolor{darkblue}{rgb}{0,0,.65}
\definecolor{darkgreen}{rgb}{0.28,0.41,0.19}

\usepackage{nicefrac}
\usepackage[%
    pdfauthor={Imre Hagymasi},%
  pdfstartview=FitH,%
  breaklinks=true,%
  bookmarks=true,%
  colorlinks=true,%
  anchorcolor=black,%
  citecolor=blue,
  filecolor=black,%
  menucolor=black,%
  urlcolor=darkblue,%
  linkcolor=blue,%
 ]{hyperref}
\usepackage[all]{hypcap} %

\newcommand{\cf}{\textit{cf.} }

\graphicspath{{images/}}
\begin{document}

\title{Enhanced symmetry-breaking tendencies in the $S=1$ pyrochlore antiferromagnet}

\author{Imre Hagym\'asi}
\affiliation{Helmholtz-Zentrum Berlin f\"ur Materialien und Energie, Hahn-Meitner Platz 1, 14109 Berlin, Germany}
\affiliation{Dahlem Center for Complex Quantum Systems and Fachbereich Physik, Freie Universit\"at Berlin, 14195 Berlin, Germany}
\affiliation{Strongly Correlated Systems "Lend\"ulet" Research Group, Institute 
for Solid State
Physics and Optics, Wigner Research Centre for Physics, Budapest H-1525 P.O. 
Box 49, Hungary
}
\author{Vincent Noculak}
\affiliation{Helmholtz-Zentrum Berlin f\"ur Materialien und Energie, Hahn-Meitner Platz 1, 14109 Berlin, Germany}
\affiliation{Dahlem Center for Complex Quantum Systems and Fachbereich Physik, Freie Universit\"at Berlin, 14195 Berlin, Germany}
\author{Johannes Reuther}
\affiliation{Helmholtz-Zentrum Berlin f\"ur Materialien und Energie, Hahn-Meitner Platz 1, 14109 Berlin, Germany}
\affiliation{Dahlem Center for Complex Quantum Systems and Fachbereich Physik, Freie Universit\"at Berlin, 14195 Berlin, Germany}
\affiliation{Department of Physics and Quantum Centers in Diamond and Emerging Materials (QuCenDiEM) group, Indian Institute of Technology Madras, Chennai 600036, India}

\date{\today}

\begin{abstract}
We investigate the ground-state properties of the nearest-neighbor $S=1$ pyrochlore Heisenberg antiferromagnet using two complementary numerical methods, density-matrix renormalization group (DMRG) and pseudofermion functional renormalization group (PFFRG). Within DMRG, we are able to reliably study clusters with up to 48 spins by keeping 20 000 SU(2) states. The investigated 32-site and 48-site clusters both show indications of a robust $C_3$ rotation symmetry breaking of the ground-state spin correlations and the 48-site cluster additionally features inversion symmetry breaking. Our PFFRG analysis of various symmetry-breaking perturbations corroborates the findings of either $C_3$ or a combined $C_3$/inversion symmetry breaking. Moreover, in both methods the symmetry-breaking tendencies appear to be more pronounced than in the $S=1/2$ system.
\end{abstract}
\maketitle

\section{Introduction}
Frustrated magnets continue to be on the forefront of condensed matter research since they often realize the fascinating situation where quantum fluctuations are strong enough to suppress the onset of magnetic order even at lowest temperatures and therefore are potential hosts for quantum spin liquids. One of the prime three dimensional candidates is the Heisenberg antiferromagnet on the pyrochlore lattice -- a cubic arrangement of corner-sharing tetrahedra -- which even in the classical limit does not order magnetically due to a ground-state degeneracy that grows exponentially in system size giving rise to a classical spin liquid.\cite{villain_insulating_1979,moecha_pyro_prl} Recent research has uncovered a striking variety of phenomena that emerges out of this classical scenario, particularly, when small effects of quantum fluctuations are taken into account. For example, adding small transverse spin interactions in a pyrochlore Ising magnet -- known as {\it quantum spin ice} -- gives rise to an emergent U(1) gauge theory, effective magnetic monopoles and emergent {\it light}.\cite{Gingras_2014,ross_quantum_2011,benton_nlce_pyrochlore_qsl_2018,taillefumier_competing_2017} 

On the other hand, the ground state properties of the Heisenberg model in the extreme quantum case, $S=1/2$, remain elusive since the combination of three spatial dimensions and strong quantum fluctuations poses significant challenges for numerical methods. Yet, there has been serious progress recently both for finite\cite{lohmann_tenth-order_2014,richter_combining_2019,schafer_pyrochlore_2020,huang_spin-ice_2016,niggemann2021quantitative,derzhko_adapting_2020,singh:dmrg2022} and zero temperatures.\cite{kim_prb_2008,burnell_monopole_2009,iqbal_quantum_2019} Numerical linked-cluster\cite{rigol_nlce_kagome_2006,khatami_nlce_checkerboard_2011,  applegate_nlce_pyrochlore_exp_comp_2012,singh_nlce_tetra_pyrochlore_2012, tang_nlce_2013, hayre_nlce_pyrochlore_2013,jaubert_nlce_pyrochlore_exp_comp_2015, benton_nlce_pyrochlore_qsl_2018,benton_nlce_pyrochlore_2018, pardini_nlce_pyrochlore_qsl_2019,schafer_pyrochlore_2020} and high-temperature series expansions\cite{lohmann_tenth-order_2014,richter_combining_2019} as well as the diagrammatic Monte Carlo technique\cite{huang_spin-ice_2016} are able to reach nontrivial temperatures but cannot target the zero temperature limit directly. Various state-of-the-art techniques, including variational Monte Carlo, \cite{astrakhantsev_broken-symmetry_2021} DMRG\cite{hagymasi_prl_2021} and PFFRG methods\cite{hering2021dimerization} point towards a magnetically disordered ground state but with broken point-group symmetries. This questions the existence of a quantum spin-liquid ground state and indicates that quantum spin ice behavior may not survive in the extreme quantum limit and for isotropic Heisenberg interactions.  
\begin{figure}[t]
    \centering
    \includegraphics[width=0.49\columnwidth]{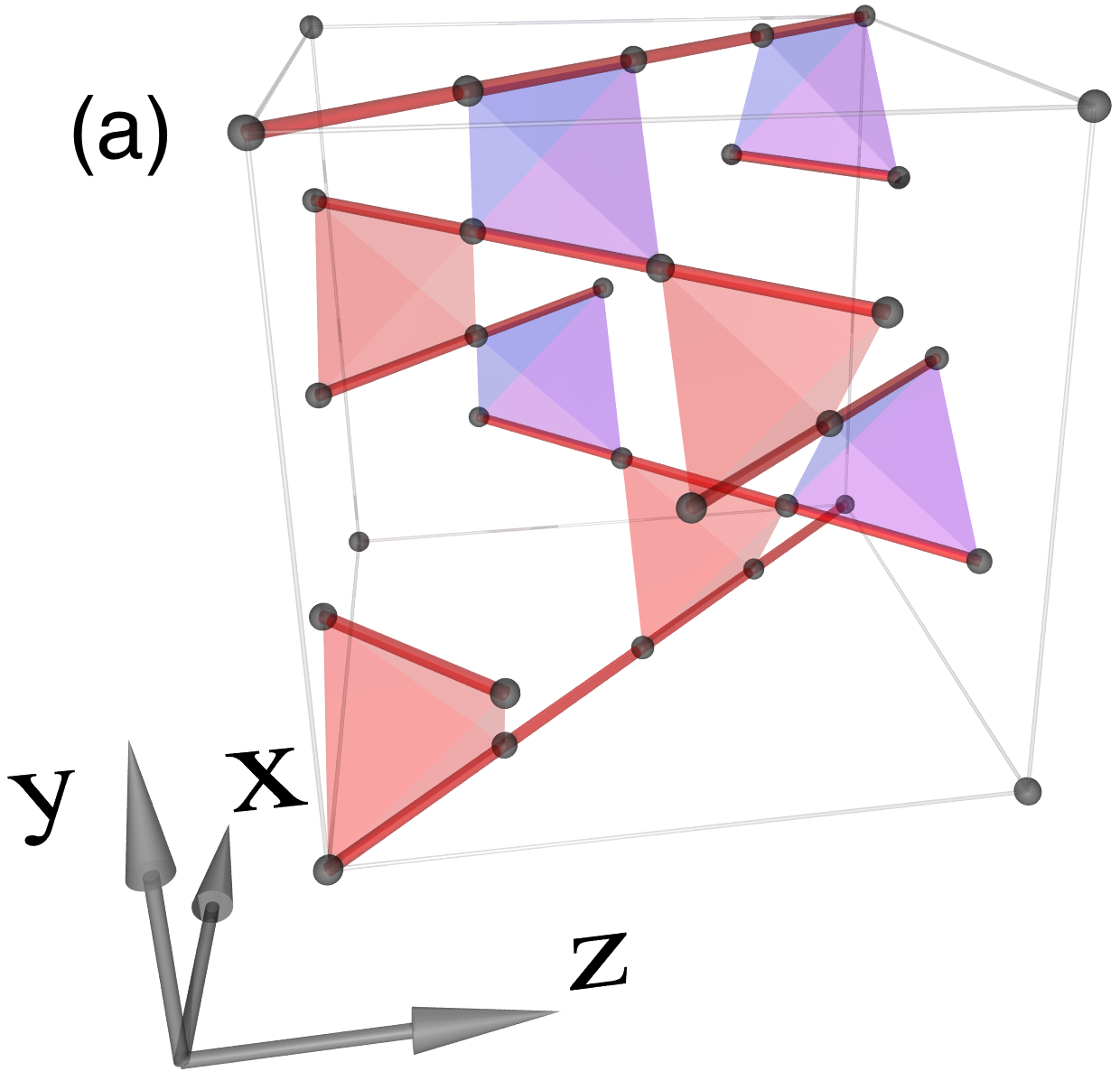}
    \includegraphics[width=0.49\columnwidth]{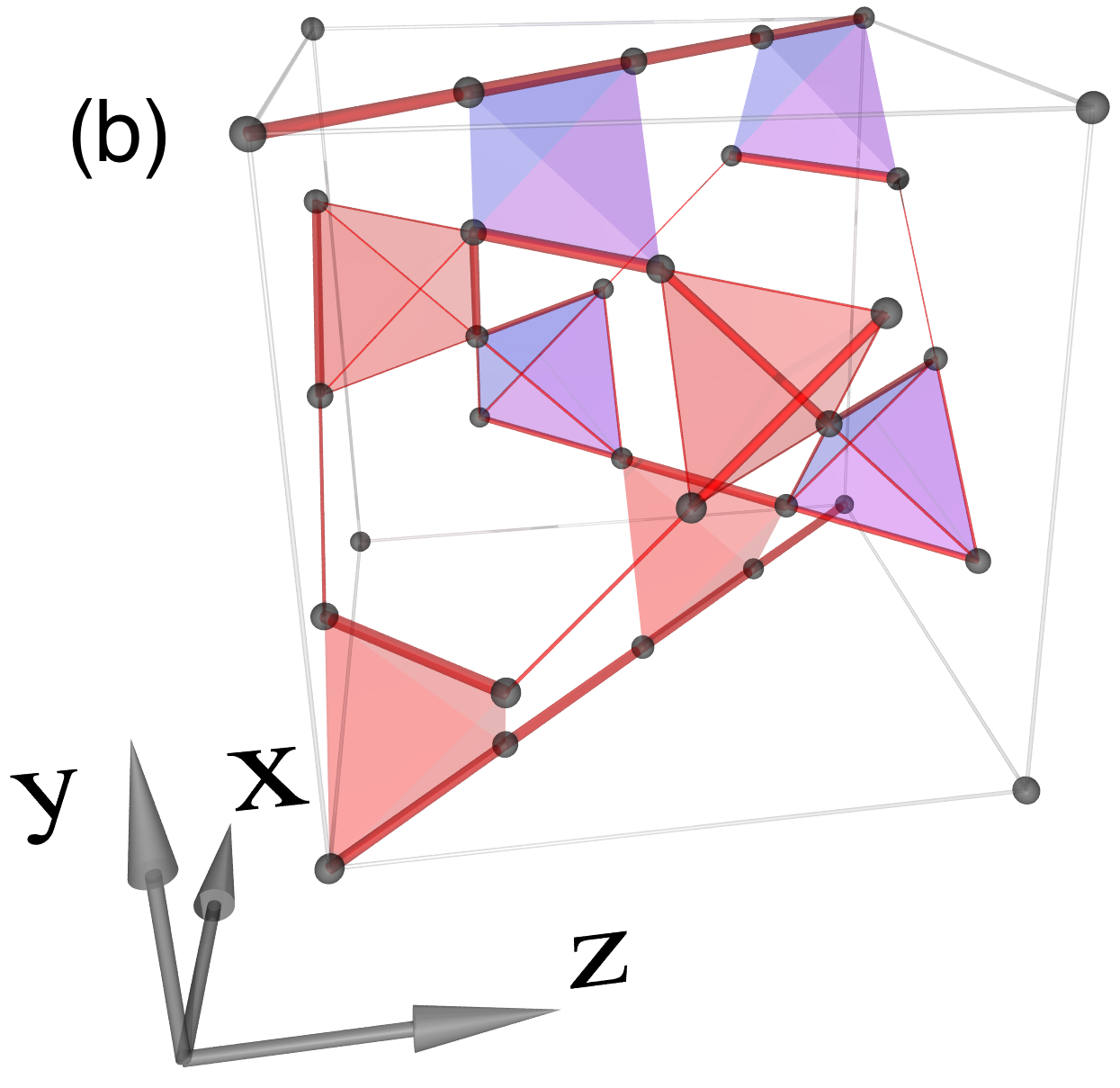}
    \caption{Nearest-neighbor spin correlations $\langle \bm{S}_i\cdot\bm{S}_j\rangle$ for the ground state of the (a) 32-site and (b) 48-site cluster. For both cases only one cubic unit cell of the clusters are shown. The widths of the lines correspond to the strength of the correlation. The shaded tetrahedral units are only guides to the eye. The values of the nearest-neighbor strong and weak spin correlations are $\langle \bm{S}_i\cdot\bm{S}_j\rangle\approx-1.36$ and $\langle \bm{S}_i\cdot\bm{S}_j\rangle\approx-0.09$, respectively for the 32-site cluster. 
}
        \label{fig:magnetization}
\end{figure}
\par In this situation, it is natural to ask what happens between the extreme quantum and the (semi-) classical limit, such as the $S=1$ case, which is even less clear than for $S=1/2$. This is due to the fact that addressing the $S=1$ case is much more difficult especially for those methods (exact diagonalization, L\'anczos techniques) that depend on the size of the Hilbert space. As an example, the current ED limit for $S=1/2$ spins is 48 sites,\cite{lauchli_kagome_2019} whose Hilbert space (in the $S^z_{\mathrm{tot}}=0$ sector) has a dimension of $\sim10^{13}$. However, for $S=1$ this Hilbert space dimension is already reached, and even exceeded, for 32 sites (which corresponds to the $2\times2\times2$ pyrochlore unit cell). Recent PFFRG studies of the $J_1$-$J_2$ model\cite{iqbal_quantum_2019} revealed that already the $S=1$ case appears to be surprisingly close to the classical limit, except for the nearest-neighbor Heisenberg model, where a nonmagnetic phase is predicted, however, its width is significantly reduced compared to the $S=1/2$ case. The rotation-invariant Green's function technique\cite{muller_thermodynamics_2019} also finds an absence of long-range magnetic ordering where the spin-spin correlation length is smaller than the lattice constant.  On another front, perturbative calculations suggest an even stronger competition between magnetically ordered and plaquette or dimer states than in the $S=1/2$ case.\cite{koga_prb_2001} In conclusion, the difference between the extreme quantum cases $S=1/2$ and $S=1$ remains elusive.
\par There is also a strong motivation to obtain new insights from experimental studies. The recently discovered NaCaNi$_2$F$_7$ compound realizes a nearly ideal $S=1$ Heisenberg model, which shows signatures of a quantum spin liquid at low temperatures. \cite{plumb_continuum_2019} 
\par In light of the very sparse results for the $S=1$ case, our goal is to make progress on the numerical front towards solving this difficult problem. With large-scale DMRG calculations we address the ground-state properties of clusters with $N=32$ and 48 sites. Our main finding is a robust $C_3$ rotation symmetry breaking and possibly an additional breaking of inversion symmetry. Remarkably, such symmetry breaking tendencies appear to be even larger compared to the spin-$1/2$ case. This is also confirmed by our complementary PFFRG analysis which indicates that either $C_3$ rotation or a combination of both $C_3$ and inversion symmetries are broken while a breaking of inversion symmetry alone seems unfavorable.

The rest of the paper is structured as follows: In Sec.~\ref{sec:methods} we introduce the model and briefly describe properties of the DMRG and PFFRG methods that are relevant for our work. Thereafter, our results from both methods are presented in Sec.~\ref{sec:results}, including real-space and momentum-space spin-correlation functions, energies of ground states and excited states as well as response functions for symmetry breaking perturbations. The paper ends with a conclusion in Sec.~\ref{sec:conclusion}.

\section{Model and methods}\label{sec:methods}
We investigate the $S=1$ Heisenberg model on the pyrochlore lattice,
\begin{equation}\label{eq_ham}
	H = J \sum_{\langle i, j\rangle} \bm{S}_i \cdot \bm{S}_j,
\end{equation}
where $\bm{S}_i=(S_i^x,S_i^y,S_i^z)^T$ is the three-component $S=1$ spin operator on site $i$.
The pyrochlore lattice is a decorated fcc lattice, with the fcc lattice vectors $\bm{a}_1 = \frac{1}{2} (1,1,0)^T$, $\bm{a}_2 = \frac{1}{2} (1,0,1)^T$, $\bm{a}_3 = \frac{1}{2} (0,1,1)^T$, together with the tetrahedral basis $\bm{b}_0=0$, $\bm{b}_i = \frac{1}{2} \bm{a}_i$, that is, each lattice point can be written as $\bm{R}_i\equiv\bm{R}_{\alpha,n_1,n_2,n_3} = n_1 \bm{a}_1 +  n_2 \bm{a}_2 +  n_3 
\bm{a}_3 + \bm{b}_\alpha,$
with integer $n_1,n_2,n_3$ and $\alpha \in \{0,1,2,3\}$.

\subsection{DMRG}
We first use the DMRG method \cite{white_1992,white_1993,noack2005,schollwock_review_2011,hallberg_review} to address the ground-state properties and low-lying excitations.  Although the method works best for one-dimensional systems, large-scale DMRG calculations have been able to give valuable results for two-dimensional\cite{white_2d_dmrg} and recently for three-dimensional systems\cite{ummethum_numerics_2013,hagymasi_prl_2021} well beyond the limitations of exact diagonalization.
However, compared to the $S=1/2$ case, the larger local Hilbert space also has an impact on the system sizes that can be treated reliably within DMRG. We consider two fully periodic clusters with $N=32$ and 48 sites. The superlattice spanned by the periodic arrangement of 32-site clusters shares the same octahedral point group $O_h$ as the fcc lattice on which the original pyrochlore lattice is based. On the other hand, this symmetry group is partially broken in the case of 48-site clusters.
The lattice vectors of the superlattice that is formed by these periodic cluster arrangements are given in Table \ref{tab:clusters}. 
\begin{table}[t]
	\centering
	\begin{tabular}{l@{\hspace{5mm}}c@{\hspace{5mm}}c@{\hspace{5mm}}c}
		\hline\hline
		cluster & $\bm{c}_1$ &  $\bm{c}_2$ & $\bm{c}_3$   \\
		\hline
		32   &  $2 \bm{a}_1$ & $2 \bm{a}_2$ & $2\bm{a}_3$   \\
		48d   &  $(1,1,1)^T$ & $(1 ,0,-1)^T$ & $(1,-1,0)^T$ \\
		\hline
	\end{tabular}
	\caption{Frame vectors $\bm{c}_1$, $\bm{c}_2$, $\bm{c}_3$ of the two clusters used in this work.  The 32-site cluster respects all point symmetries of the fcc lattice.
	The notation "d" is used in order to follow the convention of Ref. [\onlinecite{hagymasi_prl_2021}].}
	\label{tab:clusters}
\end{table}
 Enforcing the SU(2) symmetry conservation provides a much more efficient  compression for $S=1$ spins than for $S=1/2$, however, it does not compensate the growth of the Hilbert space as we increase the size of the spin. We are able to keep \mbox{20 000} SU(2) block states, which are usually equivalent to $\gtrsim 100 \ 000$ U(1) states. We map the three-dimensional cluster via a "snake" path to a one-dimensional topology, then we use the single- and two-site variants of the DMRG method\cite{hubig:_syten_toolk,hubig17:_symmet_protec_tensor_networ,hubig_2015,McCulloch_2007} to optimize the wave function and extrapolate the energies to infinite bond dimensions using the two-site variance.\cite{hubig_prb_2018} We find that the results are independent from the choice of the "snake" path and the initial state, corroborating the reliability of our calculations.

\subsection{PFFRG}
We also address the model using the PFFRG method.
This approach accesses the $T=0$ properties of a spin model via the vertex functions of the fermionic model which is obtained by mapping from spins to pseudo-fermions.\cite{Reuther10} In the standard case of spin-1/2 models, this mapping is carried out via Abrikosov's pseudo-fermion representation. To implement spin-1 degrees of freedom, we employ the approach of Ref.~[\onlinecite{Baez17}] by placing two copies of spin-1/2 operators on each site.

Within the functional renormalization group's exact infinite set of coupled differential equations, $n$-particle fermionic vertex functions are coupled to those of one order higher. By truncating this infinite hierarchy using the standard one-loop plus Katanin scheme, we obtain a finite solvable set of differential equations for the one- and two-particle vertex functions where the renormalization group parameter is an artificially introduced infrared frequency cutoff $\Lambda$ in the fermionic single-particle propagator. The resulting flow equations are then solved starting in the known infinite cutoff limit $\Lambda\rightarrow\infty$, and evolving the system towards $\Lambda\rightarrow0$ to obtain physical (i.e. cutoff-free) fermionic vertex functions. The fermionic two-particle vertex is related to the static spin-spin correlation function $\chi_{ij}^\Lambda$ which is our main numerical outcome. Necessary numerical approximations include neglecting longer-range spin-spin correlations $\chi_{ij}^\Lambda$, discretizing the vertices' frequency dependencies and applying a solving algorithm with finite cutoff step-width to the renormalization group equations. More precisely, in our calculations correlations of distances larger than five nearest-neighbor spacings are neglected. Furthermore, frequency dependencies of the two-particle vertex (self-energy) are discretized via a frequency mesh with $64$ ($2000$) points distributed exponentially around zero frequency.

Since, by construction of the PFFRG method, the spin-spin correlation functions $\chi_{ij}^\Lambda$ satisfy all symmetries of the Hamiltonian, we have to apply small bias fields to investigate whether the non-magnetic phase of the nearest-neighbor $S=1$ pyrochlore Heisenberg model tends towards spontaneous breaking of lattice symmetries. (Note that the absence of magnetic long-range order associated with broken time reversal symmetry has already been confirmed in an earlier PFFRG study.\cite{iqbal_quantum_2019}) This bias field induces a variation of the couplings $J$ in the Hamiltonian of Eq.~(\ref{eq_ham}) for different nearest-neighbor bonds $\langle i,j\rangle$, that is, $J\rightarrow J_{ij}$. More precisely, couplings are either strengthened ($J_{ij}=J+\delta$) or weakened ($J_{ij}=J-\delta$), according to the symmetry breaking pattern to be probed (here $0<\delta\ll1$).
We then monitor the system's response to such perturbations via the function $\chi_{D,ijkl}^\Lambda$ given by
\begin{equation}\label{eq:dimer_response}
    \chi_{D,ijkl}^\Lambda= \Big|\frac{J}{\delta}\frac{\chi_{ij}^\Lambda-\chi_{kl}^\Lambda}{\chi_{ij}^\Lambda+\chi_{kl}^\Lambda} \Big|\;,\text{where}\;J_{ij}=J+\delta,\;J_{kl}=J-\delta\;.
\end{equation}
Note that in the infinite cutoff limit, this response function is normalized, $\chi_{D,ijkl}^{\Lambda\rightarrow\infty}=1$. A large increase $\chi_{D,ijkl}^\Lambda\gg1$ in the cutoff free limit $\Lambda \rightarrow 0$ hints towards a phase in which the lattice symmetry, that maps the two bonds $\langle i,j \rangle$ and $\langle k,l \rangle$ onto each another, is broken.

\section{Results}\label{sec:results}
\subsection{DMRG results}
We begin with the discussion of the DMRG results for the aforementioned two types of clusters. The ground-state energies and triplet gaps are shown in Table \ref{table:energies}.
\begin{table}[t]
\centering
\begin{tabular}{@{}c|@{\hspace{4mm}}l@{\hspace{4mm}}l@{\hspace{4mm}}l@{\hspace{ 4mm}}}
\hline\hline
\centering
Cluster & GS energy & Triplet gap & Singlet gap
\tabularnewline\hline
32& $-1.5396(4)$ & 0.619(13) & 0.248(22) \tabularnewline
48d & $-1.520(6)$ &  0.51(26) & $-$ \tabularnewline
\end{tabular}
\caption{Ground-state energies per site and gaps within the $S_\text{tot}=0$
sector (singlet gap) as well as to the $S_\text{tot}=1$ sector (triplet gap) in units of $J$. The extrapolation error is defined as the half distance between the best variational energy and the extrapolated value. No value is given for the singlet gap of the 48d cluster due to large numerical costs.
}
\label{table:energies}
\end{table}
To put these numbers into context, the energy per site estimation from the rotation invariant Green's function method is $\sim-1.4J$,\cite{muller_thermodynamics_2019} which lies above our energies. Interestingly, the triplet gap of the 32-site cluster appears to be almost equal to that of the $S=1/2$ case where a triplet gap size of $0.6872J$ has been found in Ref.~[\onlinecite{hagymasi_prl_2021}]. On the other hand, for the 48d cluster the triplet gap is significantly larger in the $S=1$ case as in the spin-1/2 case where Ref.~[\onlinecite{hagymasi_prl_2021}] reports a value of $0.36(3)J$.
\par Next, we discuss the structure of the ground state by considering the nearest-neighbor spin correlations, shown in Fig.~\ref{fig:magnetization}. Surprisingly, the 32-site cluster already exhibits a well-defined pattern of spatially varying correlations. The ground state exhibits lines along which strong nearest-neighbor correlations $\langle \bm{S}_i\cdot\bm{S}_j\rangle\approx-1.36$ are present. In Fig.~\ref{fig:magnetization}(a) these lines run along the directions of the two face diagonals in the $x$-$z$ plane and do not intersect each other. All other nearest-neighbor spin correlations (i.e., those on bonds with a finite separation along the $y$-axis) are significantly weaker ($\langle\bm{S}_i\cdot\bm{S}_j\rangle\approx-0.09$), indicating that the system undergoes an effective dimensional reduction towards 1D chains. This is further supported by the fact that the strong correlations along such chains are surprisingly close to the nearest-neighbor correlations of a spin-1 Heisenberg chain, where the literature reports $\langle \bm{S}_i\cdot\bm{S}_j\rangle\approx-1.401$.~\cite{white93,todo01} The selection of the $x$-$z$ plane in this state clearly indicates a broken $C_3$ rotational symmetry around the $[111]$ axis. It is worth noting, that the same cluster with $S=1/2$ spins does not show any kind of symmetry breaking.\cite{hagymasi_prl_2021} 

\begin{figure}[t]
    \centering
    \includegraphics[width=\columnwidth]{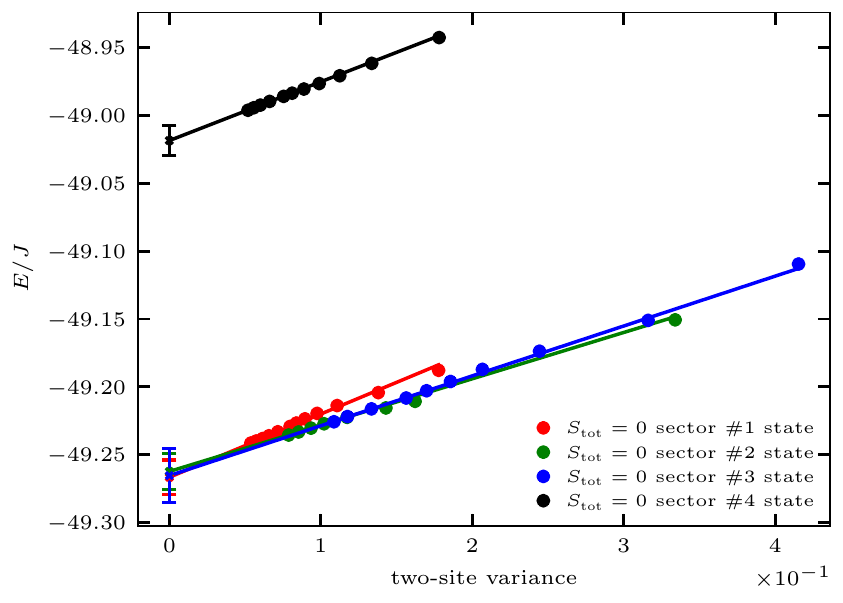}\\[0.3cm]
    \includegraphics[width=\columnwidth]{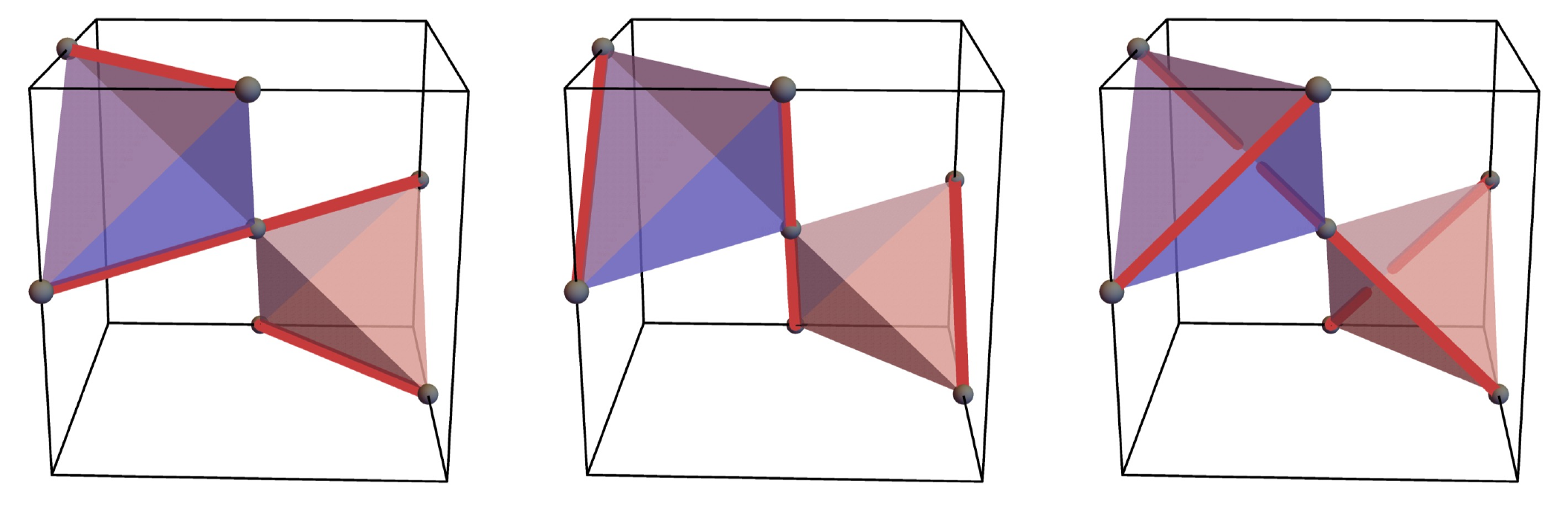}
    \caption{Extrapolation of the energies (in units of $J$) of the first four low-lying states in the $S_{\text{tot}}=0$ sector of the 32 site cluster. The data points with largest and smallest two-site variances belong to the bond dimensions 4000 and 20000, respectively.  The bottom panel shows a sketch of the three degenerate ground-state configurations, labeled by $\#1,\#2,\#3$. }
        \label{fig:orth}
\end{figure}

To put the $C_3$ rotation symmetry breaking on an even stronger footing and remove any possible bias towards low-entanglement states in the DMRG calculation, we perform an additional analysis where we identify the symmetry related states. The ground-state manifold of the 32-site cluster is expected to be threefold degenerate -- corresponding to the three choices of putting non-intersecting lines in either the $x$-$y$, $x$-$z$, or $y$-$z$ planes -- if the symmetry breaking is intrinsic. To prove that, we determine the low-lying states in the $S_{\text{tot}}=0$ sector, which is done by optimizing the wave function with the additional constraint that it should be orthogonal to the previously optimized states. (By default, DMRG converges to the state that has the smallest entanglement with respect to the snake path used.) After that, we again extrapolate the energies to the error-free limit, which is demonstrated in Fig.~\ref{fig:orth}.
Note that this approach is different from the one used in Ref.~[\onlinecite{hagymasi_prl_2021}], since it does not require additional perturbations in the Hamiltonian, however, it is only feasible for smaller system sizes due to the consecutive orthogonalization, therefore this is not feasible for the 48-site cluster. The results from Fig.~\ref{fig:orth} clearly indicate that the ground state is threefold degenerate and that the fourth level corresponds to the first singlet excitation.

\begin{figure}[t]
    \centering
    \includegraphics[width=\columnwidth]{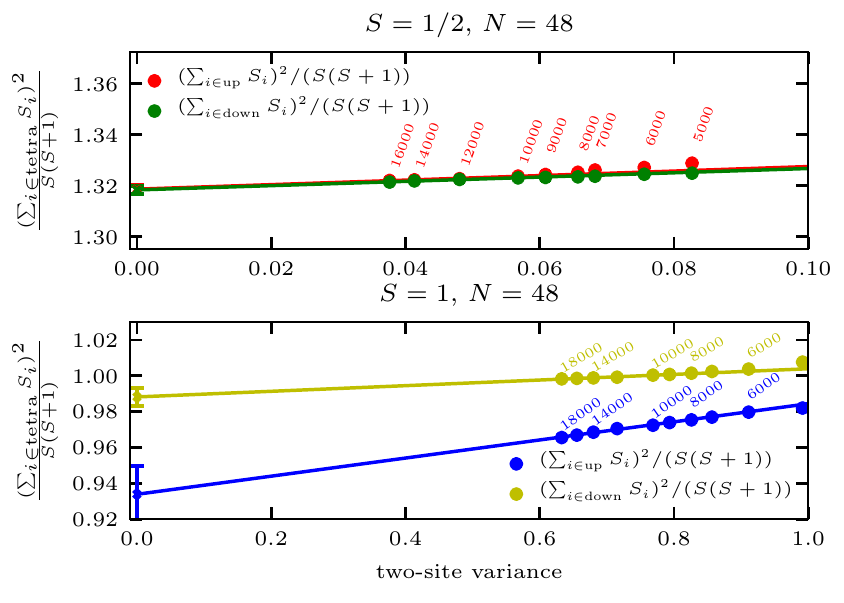}
    \caption{The (normalized) total spin squared of up and down tetrahedra for the 48d cluster. The upper and lower panels show the results for the $S=1/2$ and $S=1$ cases, respectively. The numbers above the data points denote the corresponding bond dimensions.}
        \label{fig:tetra_spin}
\end{figure}

Continuing with the 48-site cluster, the correlation pattern is more complex than for the 32-site cluster, see Fig.~\ref{fig:magnetization}(b). Although we cannot reach the same accuracy and nice convergence for the 48-site cluster like for the 32-site one (the two-site variance is 6-times larger for the 48-site cluster than for the 32-site with 20 000 SU(2) states), at first glance, the pattern looks quite similar. While some chains in the $x$-$z$ plane still show strong correlations [particularly, the ones along the $[1,0,1]$ direction in Fig.~\ref{fig:magnetization}(b)], other chains are lacking strong and homogeneous correlations. Furthermore, correlations along bonds with a finite separation along $y$-direction become relevant. 
Further details of this state are revealed when investigating the total spin of the individual tetrahedra $(\sum_{i\in\text{tetra}}\bm{S}_i)^2/(S(S+1))$, since this quantity indicates the presence or absence of the inversion symmetry. This is shown in Fig.~\ref{fig:tetra_spin}, where we compare the $S=1/2$ and $S=1$ cases for the same 48d cluster. 
In the $S=1/2$ case, the up and down energy densities merge in the error-free limit, in contrast, they converge to different values for $S=1$, which suggests that the inversion symmetry is broken for this cluster.

Altogether we can conclude that the symmetry breaking features occur already for smaller cluster sizes in the $S=1$ system compared to the $S=1/2$ case (where the $C_3$ rotational symmetry is broken at $N=48$ sites and inversion with $C_3$ rotational symmetry is broken for $N\geq64$ sites) indicating stronger symmetry breaking tendencies. \par
\begin{figure}[t]
    \centering
    \includegraphics[width=\columnwidth]{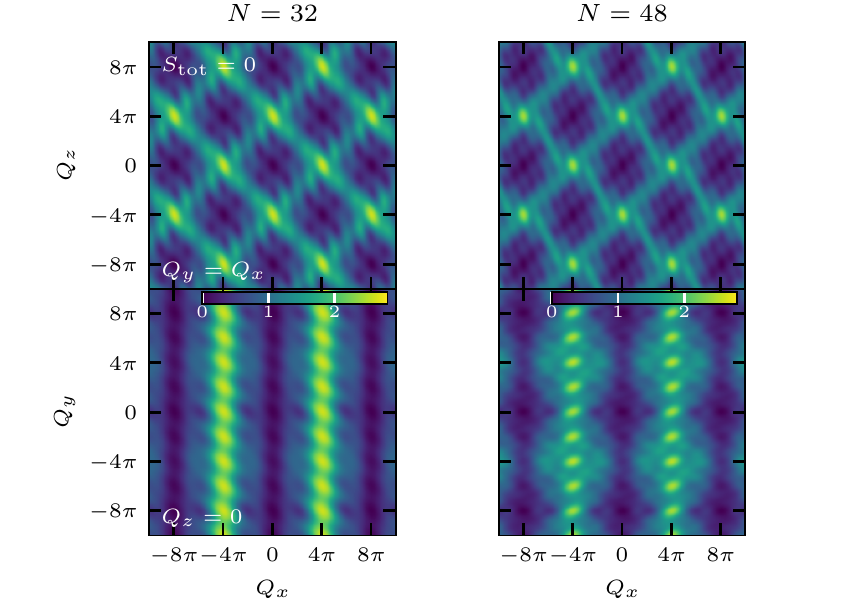}
    \caption{Spin structure factor $S(\bm{Q})$ for the two clusters. The top row shows the $Q_y=Q_x$, $[hhl]$, cut through the Brillouin zone, the bottom row corresponds to the $Q_z=0$, $[hl0]$, cut.}
        \label{fig:structure_factor}
\end{figure}

 We also calculate the equal-time spin structure factor 
\begin{equation}
    S(\bm{Q})= \frac{1}{2N} \sum_{i j} \langle \bm{S}_i\cdot 
\bm{S}_j\rangle_c \cos\left[\bm{Q}\cdot 
\left(\bm{R}_i-\bm{R}_j\right)\right],
\end{equation}
where $\bm{R}_i$ denote the real-space coordinates of sites and the index $c$ 
denotes the connected part of the correlation matrix [the factor $1/2$ comes from normalization 
$1/(S(S+1))$ for spin $S=1$]. This quantity is plotted in 
Fig.~\ref{fig:structure_factor} for the two clusters. The $[hhl]$ cut is qualitatively similar to that of the $S=1/2$ system, that is, no sharp Bragg peaks are present -- at least for these system sizes -- indicating the absence of magnetic order. The streaks in the $Q_z=0$ cut of the spin structure factor running along one selected direction clearly reflect the breaking of the $C_3$ rotational symmetry for the 32-site and 48-sites clusters. Particularly, for the 32-site cluster the signal shows almost no modulation along $Q_y$ which corresponds to the absence of correlations in the $y$-direction. For the 48-site cluster, however, modulations become more pronounced, in agreement with the onset of $y$-correlations as discussed above. 

It is worth highlighting experimental measurements of the equal-time structure factor of NaCaNi$_2$F$_7$ which is believed to be a nearly ideal realization of the nearest-neighbor $S=1$ Heisenberg antiferromagnet on the pyrochlore lattice.\cite{plumb_continuum_2019}  The  $Q_z=0$ cut in the experiments, however, does not reflect this symmetry breaking, which may be due to the fact that different domains are formed in the sample and their superpositions result in a symmetric $Q_z=0$ signal.

\begin{figure}[t]
    \centering
    \includegraphics[width=\columnwidth]{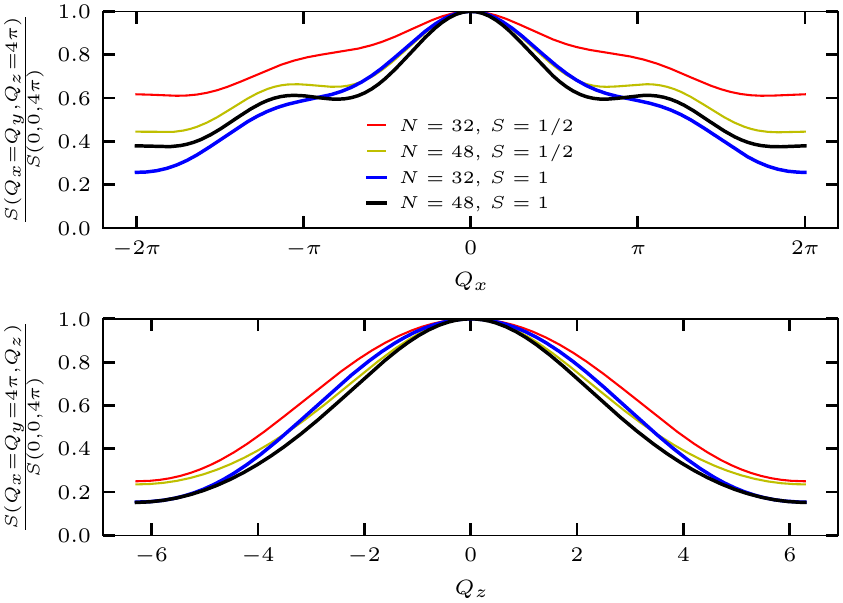}
    \caption{One-dimensional cuts of the spin structure factor across the pinch point location for the 32 and 48-site clusters as well as $S=1/2$ and $S=1$. }
        \label{fig:structure_factor_cut}
\end{figure}

Next, we discuss the fate of the pinch point feature in the spin structure factor, which is characteristic for the corresponding classical model and occurs at wave vectors ${\bm Q}=(0,0,4\pi)$ and symmetry-related points. In Fig.~\ref{fig:structure_factor_cut} we plot the spin structure factor along two one-dimensional, orthogonal cuts through this pinch point location for both the 32 and 48-site clusters as well as $S=1/2$ and $S=1$. Note that for an ideal pinch point the signal should feature a plateau around $Q_x=0$ in the top panel of Fig.~\ref{fig:structure_factor_cut} and a narrow peak at $Q_z=0$ in the bottom panel of Fig.~\ref{fig:structure_factor_cut}. All curves show a clear deviation from this ideal shape indicating that the classical pinch points are significantly altered in the $S=1/2$ and $S=1$ models and rather appear as broad peaks. Although sharp features are absent in our spin structure factors for both clusters and spin lengths, we cannot ultimately exclude the possibility that these peaks are the finite-size remnants of magnetic long-range order occurring in the thermodynamic limit. However, it is worth noting that previous studies using PFFRG\cite{iqbal_quantum_2019} and rotation-invariant Green's function method\cite{muller_thermodynamics_2019} do not detect magnetic order for $S=1$.

While we cannot perform a reasonable finite-size scaling based on our data, we find for both cluster sizes a slight narrowing of the peak along the $Q_z$-cut (bottom panel of Fig.~\ref{fig:structure_factor_cut}) as one increases the spin from $S=1/2$ to $S=1$. This is in qualitatively agreement with PFFRG\cite{iqbal_quantum_2019} and rotation-invariant Green's function results\cite{muller_thermodynamics_2019} and has been interpreted as a sign of restoration of sharp pinch points upon approaching the classical limit. Since the width of the pinch point is associated with the fulfillment of the ``ice-rule constraint'' (vanishing net magnetic moment in a tetrahedral unit) a sharpening of the pinch point should come along with a decreasing total spin per tetrahedron. This is, indeed, reflected in our results where the net spin of a tetrahedron (\cf Fig.~\ref{fig:tetra_spin}), taking into account the normalization with respect to the spin length, decreases from $\sim1.23$ for $S=1/2$ to $\sim0.92$ for $S=1$ (32-site cluster).

\subsection{PFFRG results}
\begin{figure}[t]
    \centering
    \begin{overpic}[width=\columnwidth]{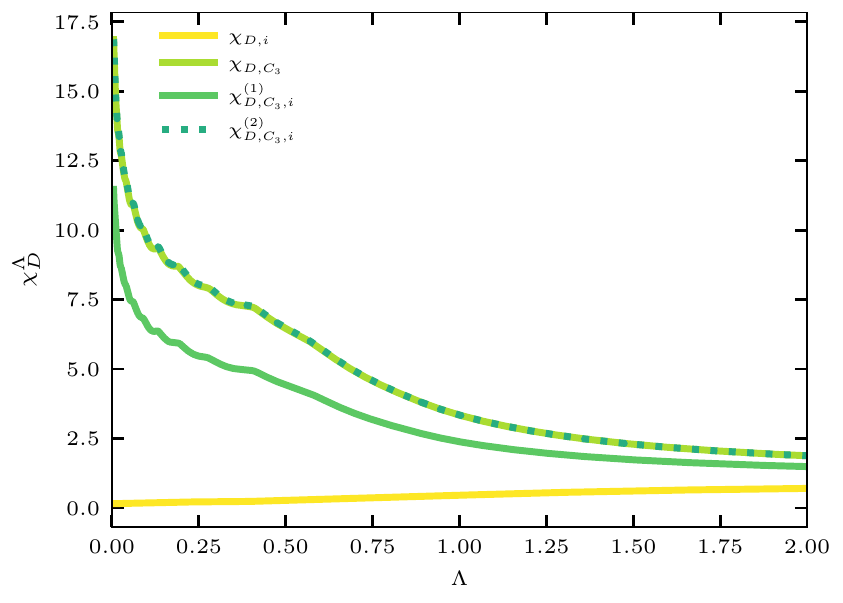}
    \put(45,27){\includegraphics[scale=0.22]{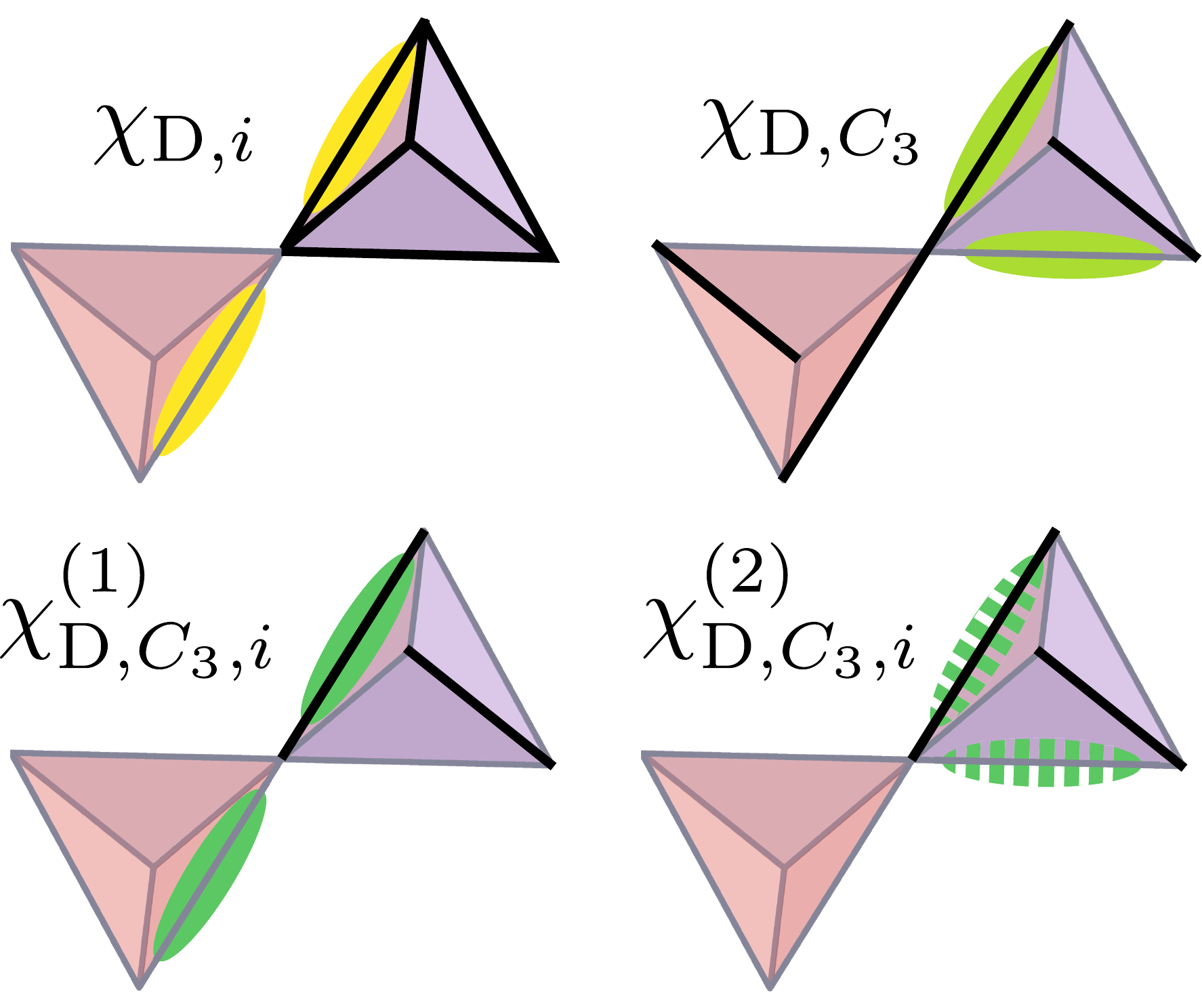}}
    \end{overpic}
    \caption{PFFRG flows of dimer responses $\chi_D^{\Lambda}$ [see Eq.~(\ref{eq:dimer_response})] for different symmetry breaking perturbations. The insets
illustrate the symmetry breaking patterns where thick black
(thin gray) lines are the strengthened (weakened) bonds with
$J \rightarrow J + \delta$ $(J \rightarrow J - \delta)$. Furthermore, the two colored bonds in
each of the four insets indicates the two bonds $(i, j)$ and $(k, l)$
for which the dimer response function in Eq.~(\ref{eq:dimer_response}) is calculated.
The response function $\chi_{D,C_3}$ probes the system with respect
to the correlation pattern in Fig.~\ref{fig:magnetization}(a). Note that $\chi^{(1)}_{D,C_3,i}$ and
$\chi^{(2)}_{D,C_3,i}$ correspond to the same perturbation but differ in the 
two bonds which are used to calculate Eq.~(\ref{eq:dimer_response}).}
        \label{fig:pffrg_flow}
\end{figure}

We will now show that, complementary to the DMRG, the PFFRG supports the picture of an enhanced lattice symmetry breaking for the ground state of the $S=1$ model compared to the $S=1/2$ model. To this end, the flow behaviors of dimer response functions $\chi_D^\Lambda$ [see Eq.~(\ref{eq:dimer_response})] for three different symmetry breaking scenarios are investigated. In a previous work~\cite{hering2021dimerization}, the same dimer response functions were studied for the $S=1/2$ model. Since otherwise the same numerical settings were used, a direct comparison is possible.

Our first perturbation (top left in the inset of Fig.~\ref{fig:pffrg_flow}) strengthens the up-tetrahedra and weakens the down-tetrahedra, resulting in a broken inversion symmetry. The corresponding response function $\chi_{D,i}$ compares two bonds related by inversion symmetry. The second pattern (top right in the inset of Fig.~\ref{fig:pffrg_flow}) is the one of Fig.~\ref{fig:magnetization}(a) where non-intersecting lines are strengthened such that $C_3$ rotation symmetry is broken. Accordingly, the response function $\chi_{D,C_{3}}$ considers two bonds which are related by a $C_3$ rotation. The last perturbation (bottom left and right in the inset of Fig.~\ref{fig:pffrg_flow}) breaks inversion and $C_3$ symmetry and corresponds to a proper dimer pattern where each site is attached to exactly one strengthened bond. In this case, two dimer responses can be defined: The bonds contributing to $\chi^{(1)}_{D,C_{3},i}$ are related by lattice inversion symmetry while those of $\chi^{(2)}_{D,C_{3},i}$ are related by lattice $C_{3}$ symmetry.

The four response functions of the $S=1$ pyrochlore Heisenberg antiferromagnet as a function of the renormalization group parameter $\Lambda$ are shown in the main panel of Fig. \ref{fig:pffrg_flow}. The results are qualitatively similar to those obtained for the $S=1/2$ system in Ref.~[\onlinecite{hering2021dimerization}].
A decrease of $\chi_{D,i}$ as the system flows from a high cutoff $\Lambda$ towards the physically relevant cutoff-free limit $\Lambda \rightarrow 0$ signifies that the ground state does not support a pure inversion symmetry breaking. On the other hand, the responses in the presence of a pure $C_{3}$ or combined $C_{3}$/inversion symmetry breaking perturbation all undergo an increase to values larger than $10$ in the cutoff-free limit indicating a strong tendency to realize these correlation patterns.
Notably, though they are not exactly equal, the largest dimer responses for those perturbations are found to be defined on bonds related by $C_{3}$ lattice symmetry, namely $\chi_{D,C_{3}}$ and $\chi^{(2)}_{D,C_{3},i}$.
Most importantly, in the limit $\Lambda \rightarrow 0$ these response functions are approximately $1.4$ times larger compared to the $S=1/2$ model which indicates enhanced symmetry-breaking tendencies in the $S=1$ case.

Since the largest response functions $\chi_{D,C_{3}}$ and $\chi^{(2)}_{D,C_{3},i}$ are almost equal, the PFFRG cannot ultimately identify the preferred symmetry breaking pattern and leaves us with the conclusion that either $C_{3}$ or combined $C_{3}$/inversion symmetries are broken, while an inversion symmetry breaking alone seems to be ruled out. This overall picture is remarkably similar to our DMRG results where, depending on the considered cluster, either a $C_{3}$ or a combined $C_{3}$/inversion symmetry breaking is found, while correlation patterns with broken inversion symmetry only are not observed.

\section{Conclusions}\label{sec:conclusion}
We investigated the ground-state properties of the $S=1$ pyrochlore Heisenberg antiferromagnet using DMRG and PFFRG. Both methods are conceptionally very different, e.g., within DMRG we study two finite and periodic spin clusters with 32 and 48 sites. On the other hand, PFFRG does not operate on a finite spin cluster but instead limits the range of spin correlations and relies on a fermionic mapping of the original spin Hamiltonian. Despite these fundamental differences, both approaches agree in their overall conclusion, in that the tendency towards rotational or combined rotational and inversion symmetry breaking is stronger in the $S=1$ case than for the $S=1/2$ system. The inherent limitations of both methods should, however, also be discussed. For example, the small cluster sizes studied with DMRG do not allow an extrapolation to the thermodynamic limit. While a possible cluster bias does not exist within PFFRG, this approach relies on approximated (truncated) renormalization group equations where multispin-correlations are poorly incorporated. Hence, despite the coherent physical scenario revealed in this paper, we believe that further investigation of this difficult problem is necessary.

It is again worth putting these results in the context of available experimental findings.\cite{plumb_continuum_2019}  While our results for the spin structure factor are consistent with experimental data, a symmetry breaking in the ground state must be accompanied by a finite temperature phase transition, which could serve as a corroboration of our findings. The reported specific heat and entropy data down to $100$ mK do not exhibit any anomalies; it would, hence, be interesting to search for a possible phase transition at even lower temperatures.

\begin{acknowledgments}
We thank Robin Sch\"afer for helpful discussions.
I.\,H.~was supported in part by the Hungarian National Research,   
Development   and   Innovation Office (NKFIH) through Grants No.~K120569 and No.~K134983.
V.\,N. and I.\,H.  would like to thank the HPC service of ZEDAT, Freie Universität Berlin, and the Department of Physics, Freie Universität Berlin, for computing time at the Curta and Tron cluster. Some of the data presented here was produced using the \textsc{SyTen} toolkit. \cite{hubig:_syten_toolk,hubig17:_symmet_protec_tensor_networ} J.\,R. thanks IIT Madras for a Visiting Faculty Fellow position under the IoE program during which part of the research work was carried out.
\end{acknowledgments}

\bibliography{pyrochlore,pyrochlore_field,pyrochlore2}
\clearpage
\appendix

\end{document}